\documentclass[12pt,preprint]{aastex}

\newcommand{\lsim}{\raise0.3ex\hbox{$<$}\kern-0.75em{\lower0.65ex\hbox{$\sim$}}}
\newcommand{\gsim}{\raise0.3ex\hbox{$>$}\kern-0.75em{\lower0.65ex\hbox{$\sim$}}}

\newcommand{\kms}{\rm ~km~s^{-1}}
\newcommand{\ergs}{\rm ~erg~s^{-1}}

\newcommand{\wl}{\lambda}

\newcommand{\ml}{~M_\odot ~\rm yr^{-1}}

\def\EE#1{\times 10^{#1}}

\newcommand{\Mdot}{\dot M}

\begin{document}

\title{The X-ray and radio emission from SN 2002ap: \\
The importance of Compton scattering}

\author{
Claes-Ingvar Bj\"ornsson\altaffilmark{1},
Claes Fransson\altaffilmark{1},
}

\altaffiltext{1}{Department of Astronomy, AlbaNova University Center, SE--106~91 Stockholm,
Sweden.} 
\email{}

\begin{abstract}
The radio and X-ray observations of the Type Ic supernova SN 2002ap
are modeled. We find that inverse Compton cooling by photospheric
photons explains the observed steep radio spectrum, and also the X-ray
flux observed by XMM. Thermal emission from the shock is insufficient
to explain the X-ray flux. The radio emitting region expands
with a velocity of $\sim 70,000 \kms$.  From the ratio of X-ray to
radio emission we find that the energy densities of magnetic fields
and relativistic electrons are close to equipartion. The mass loss 
rate of the progenitor star depends on the
absolute value of $\epsilon_{\rm B}$, and is given by $\Mdot \approx
1\EE{-8} (v_{\rm w}/1000 \kms)\epsilon_{\rm B}^{-1} \ml$.  
\end{abstract}


\section{Introduction}
The connection between Type Ic supernovae (SNe) and gamma-ray bursts (GRBs), as
recently highlighted by GRB030329=SN2003dh \citep{Stan03,Hjo03,Kaw03},
have made a detailed understanding of this class of SNe especially
urgent. The first indication of an association of GRBs with Type Ic 
SNe was, however,
already clear from the observations of GRB980425=SN1998bw
\cite[e.g.,][]{Iwa03}. Type Ic SNe are believed to be the result of explosions
of Wolf-Rayet stars, which have lost their hydrogen and most of their
helium envelopes \citep{N94,WLW95}. Progenitors of this type have strong
stellar winds with mass loss rates $\Mdot \sim 10^{-5} \ml$ and wind
velocities $v_{\rm w} \sim 1000-4000\kms$. The interaction of the
non-relativistic supernova ejecta with this wind has long been
invoked to explain the radio and X-ray emission observed for this
type of SNe \citep[e.g.,][]{CF03}. For the GRBs a relativistic
version of the same scenario has been used to explain the
properties of the GRB afterglows \citep[e.g.,][]{LC03}. 

Because of its small distance, 7.3 Mpc, SN 2002ap has been subject to
observations in several wavelength bands from radio to X-rays. The
optical spectrum showed the SN to be a typical Type Ic, with no
evidence for hydrogen or helium. Although line blending made it hard 
to determine velocities accurately, 
\citet{Maz02} argued that SN 2002ap belongs to the extremely energetic events sometimes
referred to as 'hypernovae', where the prime example is SN 1998bw.
The total kinetic energy is, however, sensitive to the fraction of the
ejecta at very high velocity, which is not probed by the optical
spectra. Radio and X-ray observations are more sensitive to this part
of the ejecta, since radiation in these wavelength bands are expected 
to arise in the shocked, faster moving outer material. 
Hence, an accurate modeling of the radio and X-ray
observations is well motivated.

Radio observations of SN 2002ap with VLA, and modeling of these, have
been discussed in \citet{BKC02} (in the following BKC). With the 
assumption of equipartion between magnetic fields and relativistic 
electrons, these authors
showed that the radio light curves could be explained by a
synchrotron self-absorption model, where the radio emitting material
expands at $\sim 0.3$ c. While the fits to the light curves were
satisfactory, there are some issues left open by the model. In
particular, the implied optical thin synchrotron spectrum is 
considerably flatter than the observed one. The steeper than expected 
spectrum could be due to synchrotron cooling. If so, a magnetic field 
strength much in excess of equipartion is required.

In parallel to this, there has been several papers discussing XMM
observations of SN 2002ap \citep{SK02,SCB03,SPM03}. These authors have 
interpreted the X-ray emission as coming from the thermal electrons 
behind the shock either directly as free-free emission or as inverse 
Compton scattering of the photospheric SN photons. For such models to work, 
the shock velocity has to
be fairly low ($\sim 10,000-20,000\kms$ at day 6). As we argue in 
this paper, there is evience from optical line profiles that velocities higher than 
this are needed. Furthermore, a low shock velocity implies a small 
size of the radio emission region and, hence, a high brightness 
temperature.

In this paper we show that a consistent picture of both the radio and
X-ray observations can be obtained by a model where inverse Compton
cooling of the relativistic electrons by the photospheric photons is
important. This explains both the steep radio spectrum, and the X-ray
emission as coming from the same region, expanding at a velocity
compatible with both optical and radio observations. The inclusion of
the X-ray constraint also allows a determination of the 
ratio between the energy densities in magnetic fields and relativistic electrons.

The paper is organized as follows. In \S \ref{sec_2} we discuss the
constraints on different cooling processes of the relativistic
electrons, using simple analytical arguments. In \S \ref{sec_3} we
illustrate this by a detailed model calculation of the observed light
curves and radio and X-ray spectra. The implications of this and a
comparison with related papers are given in \S \ref{sec_disc}.  In \S
\ref{sec_concl} we summarize our conclusions. We will in the following
use a distance of $D\approx 7.3$ Mpc for M 74 \citep{SKT96}.

\section{Analytical considerations.}
\label{sec_2}
An approximate expression for the optically thin synchrotron
luminosity, $L_{\nu}$, emitted at a frequency $\nu$ from behind a
spherically symmetric, non-relativistic shock with velocity $v_{\rm
sh}$ and radius $R_{\rm sh}$ is given by
\begin{equation}
     \nu L_{\nu}\approx
     \pi R_{\rm sh}^2 
v_{\rm sh}n_{\rm rel}\left(\frac{\gamma_{\nu}}{\gamma_{\rm min}}\right)^{1-p}
     \gamma_{\nu}mc^2
     \left(1+\frac{t_{\rm synch}(\nu)}{t}+\frac{t_{\rm synch}(\nu)}{t_{\rm other}
     (\nu)}\right)^{-1}.
     \label{eq:1.1}
\end{equation}
\citep[e.g.,][]{FB98}. The number density of
relativistic electrons is denoted by $n_{\rm rel}$ and $\gamma_{\nu}$ is
the Lorentz factor of electrons radiating at a typical frequency
$\nu \propto\gamma_{\nu}^2 B$, where $B$ is the strength of the
magnetic field. The energy distribution of the relativistic
electrons injected behind the shock is parameterized as
$dn_{\rm rel}(\gamma)/d\gamma  \propto\gamma^{-p}$ for
$\gamma_{\rm min}\leq\gamma\leq\gamma_{\rm max}$ and zero otherwise. 
The second bracket on the RHS of equation
(\ref{eq:1.1}) is the fraction of the injected energy which is
radiated away as synchrotron radiation. This is determined by three
different time scales: $t$ is the dynamical time (i.e., the adiabatic
cooling time; roughly, the time since the onset of expansion),
$t_{\rm synch}(\nu)$ is the synchrotron cooling time for electrons
radiating at a frequency $\nu$, and $t_{\rm other}(\nu)$ is the cooling
time corresponding to processes other than synchrotron
radiation.

The extensive observations of SN 1993J allowed a detailed
modeling of the synchrotron radiation produced by the shock. It was
shown in \citet{FB98} that behind the shock, the energy densities of
relativistic electrons and magnetic field both scaled with the thermal
energy density. Furthermore, the shock expanded into a circumstellar
medium in which density ($n_{\rm csm}$) varied with radius as $R^{-2}$. As a result,
$n_{\rm rel}\propto B^2\propto n_{\rm csm}v_{\rm sh}^2\propto t^{-2}$. Implicit in these
scaling relations is the assumption of no time variation in $\gamma_{\rm min}$.
However, since the deduced value of $p$ was close to two, any variation
in $\gamma_{\rm min}$ (and/or $\gamma_{\rm max}$) would only introduce a 
logarithmic correction.
The radio observations of SN 2002ap are such that a detailed
modeling is not possible. Hence, in order to limit the scope of our
discussion, it will be assumed that the scaling relations obtained
for SN 1993J are applicable also for SN 2002ap. This is not a 
critical assumption; for example, as is shown below, the different scaling relations 
used by BKC give much the same results. With this background
we will now discuss two alternative scenarios, where synchrotron
respectively inverse Compton cooling are responsible for the steep
radio spectrum. 

\subsection{Synchrotron radiation as the dominant cooling process}
\label{sec_2.1}
Consider first the case when synchrotron radiation dominates other
cooling processes (i.e., $t_{\rm synch}<t_{\rm other}$). With the use of
$R_{\rm sh}\approx v_{\rm sh}t$, one obtains from equation (\ref{eq:1.1}) the
time variation of the optically thin synchrotron luminosity
\begin{equation}
     L_{\nu}\propto v_{\rm sh}^3 \gamma_{\nu}^{2-p}\left(1+\frac{t_{\rm synch}}{t}
     \right)^{-1}.
     \label{eq:1.2}
\end{equation}
When radiative cooling is {\it not} important (i.e., $t<t_{\rm synch}$), this
leads to
\begin{eqnarray}
     L_{\nu}&\propto& v_{\rm sh}^3 B^{(p+1)/2}t\nonumber\\
     &\propto& v_{\rm sh}^3 t^{(1-p)/2},
     \label{eq:1.3}
\end{eqnarray}
where $t_{\rm synch}\propto B^{-3/2}$ has been used. When synchrotron
cooling  {\it is} important (i.e., $t>t_{\rm synch}$), the
corresponding expression for the synchrotron luminosity is
\begin{eqnarray}
     L_{\nu}&\propto& v_{\rm sh}^3 B^{(p-2)/2}\nonumber\\
     &\propto& v_{\rm sh}^3 t^{(2-p)/2}.
     \label{eq:1.4}
\end{eqnarray}

The time dependence of $v_{\rm sh}$ in SN 2002ap used by BKC is the one
for an ejecta whose structure is thought appropriate for a Ic SN, 
expanding into a $n_{\rm csm}\propto R^{-2}$ circumstellar medium.  The
self-similar solution \citep{RAC82} then gives $v_{\rm sh}\propto 
t^{-1/(n-2)}$, where $n$ is the effective power law index of the ejecta 
density. For the high velocities relevant for SN 2002ap,  \citet{MMK99}
find $n=10.18$ or $v_{\rm sh}\propto t^{-0.12}$. 
The observed optically thin synchrotron flux from SN
2002ap varies, roughly, as $F_{\nu}\propto t^{-0.8}$ (BKC). It is seen from
equation (\ref{eq:1.3}) that this time dependence can be obtained for
$v_{\rm sh}\propto t^{-0.12}$ and $p\approx 2$. This is the same
conclusion as that reached by BKC, although they used somewhat
different scaling relations for the energy densities behind the shock.
It is also consistent with an equipartition value for the magnetic
field, since such a value implies $t<t_{\rm synch}$. The latter is a
necessary requirement for the validity of equation
(\ref{eq:1.3}). However, this solution predicts an optically thin
spectral index $\beta\approx 0.5$, where $L_{\nu}\propto
\nu^{-\beta}$. This is not consistent with the observed value
$\beta\approx 0.9$. This discrepancy is exacerbated by the fact that
the spectral range used to determine $\beta$ is rather close to where
optical depth effects become important. Hence, spectral curvature
could make the actual spectral index of the optically thin radiation
somewhat larger than the measured value of $\beta$.

Short term variations are apparent in the radio light curves. As 
suggested by BKC, this is likely due to interstellar 
scattering and scintillation. These variations cause substantial 
variations in the instantaneous spectral index, which indicates that they 
are not broad-band. Therefore, a spectral index measured at a given time 
should be treated with some care. The typical time scale 
for such flux variations is at most a few hours; hence, the 
ratio of the fluxes at 4.86 GHz and 8.46 GHz averaged over several 
days should give a good estimate of the intrinsic optically thin 
spectral index of SN 2002ap. The value quoted above 
for the spectral index (i.e., $\beta\approx 0.9$) was obtained by averaging over several 
observing periods and should thus be robust.

An alternative solution would be to assume $t>t_{\rm synch}$ (i.e.,
synchrotron cooling is important) and $p\approx 2$, since this gives a
value of $\beta$ consistent with the observed one. The luminosity is
then obtained from equation (\ref{eq:1.4}) as $L_{\nu}\propto v_{\rm
sh}^3$, which implies $v_{\rm sh}\propto t^{-0.27}$, or $n \approx
6$. Incidentally, this is the same time dependence of the velocity as
determined for the type IIb SN 1993J. However, this scenario has a 
few not so attractive implications, as can be seen from the following 
arguments. 

The thermal energy density ($U_{\rm th}$) behind a non-relativistic
supernova shock wave expanding into the circumstellar progenitor wind,
varies with time ($t_{d}$, in days) as
\begin{equation}
     U_{\rm th}=7.6\times 10\,\frac{\dot{M}_{-5}}{v_{\rm w,3}}t_{\rm d}^{-2} ~\rm ergs ~cm^{-3},
     \label{eq:1.5}
\end{equation}
where $\dot{M}_{-5}$ is the mass loss rate of the progenitor star in
units of $10^{-5} \ml$ and $v_{\rm w,3}$ is its wind velocity in units
of $10^3 \kms$. The synchrotron cooling time for electrons radiating
at a frequency $\nu$ is
\begin{equation}
     t_{\rm synch}= 1.7\times 10^2 B^{-3/2}\nu_{10}^{-1/2} ~\rm days,
     \label{eq:1.6}
\end{equation}
where $\nu_{10}\equiv \nu/10^{10}$. Furthermore, parameterizing the
magnetic field strength as $B^2/8\pi\equiv\epsilon_{\rm B}U_{\rm th}$, the
requirement for synchrotron cooling to be important (i.e.,
$t_{\rm synch}<t$) can be written
\begin{equation}
     t<3.0\, \epsilon_{\rm
     B}^{3/2}\nu_{10}\left(\frac{\dot{M}_{-5}}{v_{\rm w,3}} 
     \right)^{3/2}  ~\rm days.
     \label{eq:1.7}
\end{equation}

In the cooling scenario, observations indicate that cooling is
still important at $4.86$ GHz for $t=15$ days (BKC). From equation
(\ref{eq:1.7}) this implies
\begin{equation}
     \epsilon_{\rm B}\frac{\dot{M}_{-5}}{v_{\rm w,3}}>4.7.
     \label{eq:1.8}
\end{equation}
Since physically $\epsilon_{\rm B}<1$, equation (\ref{eq:1.8}) shows that
cooling due to synchrotron radiation requires a very high mass loss rate.
Under such conditions free-free absorption in the circumstellar
wind may become important. A realistic estimate of the free-free
optical depth, however, necessitates a self-consistent determination
of the temperature ahead
of the shock \citep[cf. the discussion of SN 1993J in][]{FB98}.
This is not a straightforward exercise and, in addition, the result
is rather model dependent \citep{LF88}. 

However, there is another consequence of the synchrotron cooling
scenario that argues against its applicability to SN 2002ap. If
cooling is important for electrons radiating at a frequency $\nu$,
then
\begin{equation}
     \nu L_{\nu}\approx \pi R^2 v_{\rm sh}U_{\rm th}\frac{\epsilon_{\rm rel}}{\ln
     (\gamma_{\rm max}/\gamma_{\rm min})},
     \label{eq:1.9}
\end{equation}
where $\epsilon_{\rm rel}$ is
the fraction of the thermal energy, which goes into relativistic
electrons. The logarithmic factor in
equation (\ref{eq:1.9}) results from $p=2$. With the use of the
expression for $U_{\rm th}$, equation (\ref{eq:1.9}) can be written
\begin{equation}
     \epsilon_{\rm rel}\approx \frac{32}{9}\frac{\nu L_{\nu}\,\ln(\gamma_{\rm max}/
     \gamma_{\rm min})}{c^3}\,\frac{v_{\rm w}}{\dot{M}}\,\left(\frac{c}{v_{\rm sh}}
     \right)^{3}.
     \label{eq:1.10}
\end{equation}
The constraint in equation (\ref{eq:1.8}) then yields
\begin{equation}
     \frac{\epsilon_{\rm rel}}{\epsilon_{\rm B}}< 1.5\times 10^{-8}\ln(\gamma_{\rm max}/
     \gamma_{\rm min})\left(\frac{c}{3v_{\rm sh}(t_{\rm d}=5)}\right)^{3},
     \label{eq:1.11}
\end{equation}
where the expression in equation (\ref{eq:1.10}) has been evaluated
at $t=5$ days. As discussed in \S \ref{sec_disc}, the widths of 
the observed optical emission lines indicate
velocities $\ga 0.1$ c, which is a lower
limit for $v_{\rm sh}$. Hence, equation (\ref{eq:1.11}) requires an
unprecedented low value of $\epsilon_{\rm rel}/\epsilon_{\rm B}$.
Although such a situation cannot be excluded, we discuss in the next
section a scenario in which cooling is dominated by Compton
scattering rather than synchrotron radiation.

\subsection{Compton scattering as the dominant cooling process}
\label{sec_2.2}
When Compton scattering dominates the cooling (i.e.,
$t_{\rm comp}<\mbox{min}[t,t_{\rm synch}]$, where $t_{\rm comp}$ is the time scale
for Compton cooling), the time variation of the optically thin synchrotron
radiation is obtained from equation (\ref{eq:1.1}) as
\begin{eqnarray}
     L_{\nu}&\propto&v_{\rm sh}^3\gamma_{\nu}^{2-p}\frac{t_{\rm comp}}{t_{\rm synch}}
     \nonumber\\&\propto&\frac{v_{\rm sh}^{3}B^{(p+2)/2}}{U_{\rm ph}},
     \label{eq:1.12}
\end{eqnarray}
where $U_{\rm ph}$ is the energy density of photons. Consider first the
case when cooling is due to the external photons from the supernova
itself; hence, $U_{\rm ph}=L_{\rm bol}/(4\pi R_{\rm sh}^2c)$, where $L_{\rm bol}$ is the
bolometric luminosity of the supernova. This then implies from
equation (\ref{eq:1.12})
\begin{equation}
     L_{\nu}\propto \frac{v_{\rm sh}^5 t^{(p-2)/2}}{L_{\rm bol}}.
     \label{eq:1.13}
\end{equation}
When Compton cooling is strong (i.e., $t_{\rm comp}\ll t$), the light
curves for the optically thin frequencies are likely to show a minimum
close to where the maximum of $L_{\rm bol}$ occurs. In cases where
also the optically thick-thin transition is observed, this would give
rise to a double peaked structure. The observed time variation of
$L_{\rm bol}$ is shown in Figure 1. It is seen that the bolometric
luminosity actually peaks during the main radio observing
period. Although the observations
span a rather limited range in time and, furthermore, are likely to be
affected by interstellar scattering and scintillation (BKC), it is
noteworthy that no pronounced minimum is apparent in the light
curves for the optically thin frequencies. This can be understood if $t_{\rm comp}$ is roughly equal to
$t$. Before making a more detailed numerical fit to the data, we will
show that if $t_{\rm comp}\sim t$ we can obtain a consistent and
plausible set of values for $v_{\rm sh}$, $\epsilon_{\rm B}$ and
$\epsilon_{\rm rel}$.

The time scale for Compton cooling is given by
\begin{equation}
     t_{\rm comp}=6.6\frac{B^{1/2}}{\nu_{10}^{1/2}U_{\rm ph}}  ~\rm days.
     \label{eq:1.14}
\end{equation}
With the use of
\begin{equation}
     U_{\rm ph}=4.0\times 10^{-1}\frac{L_{\rm bol,42}}{t_{d}^2}
     \left(\frac{v_{\rm sh}}{c}\right)^{-2}  ~\rm ergs ~cm^{-3},
     \label{eq:1.15}
\end{equation}
the condition $t_{\rm comp}\sim t$ results in
\begin{equation}
     L_{\rm bol,42}\sim 10\,B^{1/2}\left(\frac{3v_{\rm sh}}{c}\right)^{2},
     \label{eq:1.16}
\end{equation}
where $L_{\rm bol,42}\equiv L_{\rm bol}/10^{42} \ergs$. Furthermore,
$\nu_{10} = 1$ and $t = 6$ days have been used. As an
illustration we show in Figure \ref{fig1} the ratio of the Compton 
cooling time
scale to the adiabatic time scale for the specific model in \S \ref{sec_3} with
$\Mdot=10^{-5} \ml$, $v_{\rm w}=1000 \kms$, $v_{\rm sh}(t_{\rm d}=10)=70,000 \kms$,
and $\epsilon_{\rm B}=10^{-3}$. For other values $ t_{\rm comp}
\propto (\epsilon_{\rm B} \Mdot/v_{\rm w})^{1/4} v_{\rm sh}^2 t^{3/2}
L_{\rm bol}^{-1} \nu_{10}^{-1/2}$.

An X-ray flux was detected from SN 2002ap by XMM-Newton on 2002, Feb\,3 ($t\approx 6$ days)
\citep{SK02,SCB03,SPM03}. Due to 
the weak signal neither the total flux nor the spectral shape could be well 
determined. The uncertainty in the high energy flux is affected by 
the subtraction of the flux from a strong, nearby source with a hard 
spectrum,  while at low energy the inability to constrain absorption in 
excess of the 
galactic value makes the observed flux a lower limit to the intrinsic one. 
The observations can be fitted either with a thermal or a 
power law spectrum. In the latter case, the deduced spectral index is consistent with 
that observed in the radio. It is therefore likely
that the X-ray flux is the Compton scattered optical radiation from
the supernova. If this is the case, independent estimates of the
energy densities in magnetic fields ($U_{\rm B}$) and relativistic
electrons ($U_{\rm rel}$) can be obtained. Since $p\approx 2$
\begin{equation}
     \frac{F_{\rm radio}}{F_{\rm X}}\sim \frac{U_{\rm B}}{U_{\rm ph}},
     \label{eq:1.16a}
\end{equation}
where $F_{\rm radio}\approx \nu F_{\nu}$ is the monochromatic optically thin 
synchrotron flux and $F_{\rm X}$ is the corresponding 
X-ray flux. As it turns out, the deduced value of $B$ is
such that the same electrons, roughly, are producing both the radio
and X-ray fluxes. Hence, equation (\ref{eq:1.16a}) is not sensitive to
the exact value of $p$. Assuming no intrinsic absorption and a 
power law spectrum with a spectral index consistent with that in the 
radio give at $t\approx 6$ days a monochromatic X-ray flux $F_{\rm X}\approx
(3.0\pm0.5)\times 10^{-15} ~\rm ergs ~cm^{-2}~s^{-1}$.  At the same 
time, $F_{\rm radio}\approx 2.0
\times 10^{-17} \rm ergs~cm^{-2}~s^{-1}$ and one finds from equation
(\ref{eq:1.16a})
\begin{equation}
     U_{\rm B}\sim 6 \times
     10^{-4}L_{\rm bol,42}\left(\frac{3v_{\rm sh}}{c}\right)^{-2}  ~\rm ergs ~cm^{-3}.
     \label{eq:1.17}
\end{equation}
If now $t_{\rm comp}\sim t$,  equations (\ref{eq:1.16}) and (\ref{eq:1.17}) 
lead to $U_{\rm B}\sim 6\times 10^{-3} B^{1/2}$; hence, $B\sim 0.3~\rm 
G$ and $U_{\rm B} \sim 3\times 10^{-3}~\rm ergs ~cm^{-3}$. Together
with the observed value  $L_{\rm bol,42}\approx 1.6$ at $t\approx 6$ days,
this implies $v_{\rm sh}/c \sim 0.2$.

With a supernova distance $7.3$ Mpc, the observed X-ray flux
corresponds to a monochromatic X-ray luminosity $L_{\rm X}\approx 1.9\times 10^{37}\ergs$,
from which the energy density in relativistic electrons is obtained as (cf.
eq. [\ref{eq:1.1}])
\begin{equation}
     U_{\rm rel}\sim \frac{4~L_{\rm X} \ln (\gamma_{\rm max}/\gamma_{\rm min})}{\pi v_{\rm sh}^3
     t^2},
     \label{eq:1.18}
\end{equation}
where, again, the logarithmic factor is due to $p\approx 2.0$.
Furthermore, equation (\ref{eq:1.18}) assumes $t_{\rm comp}\sim t$ so that,
roughly, half of the injected energy emerges as X-ray flux and that,
in turn, half of this is emitted through the forward shock (the
other half being absorbed by the supernova ejecta). Hence, at
$t = 6$ days, $U_{\rm rel}\sim 1\times 10^{-4}(3v_{\rm sh}/c)^{-3}
\ln(\gamma_{\rm max}/\gamma_{\rm min})  ~\rm ergs ~cm^{-3}$. With the 
use of $v_{\rm sh}/c \sim 0.2$, this leads to $U_{\rm rel}\sim 5\times 10^{-4}
\ln(\gamma_{\rm max}/\gamma_{\rm min})  ~\rm ergs ~cm^{-3}$.

Since the unknown values of $\gamma_{\rm max}$ and $\gamma_{\rm min}$ 
enter only logarithmically in $U_{\rm rel}$, this shows that in the Compton cooling 
scenario for SN 2002ap, where the
relativistic electrons cool on the external photons from the supernova
itself, $U_{\rm B}\sim U_{\rm rel}$, i.e., there is rough equipartition between
the energy densities in magnetic fields and relativistic electrons.
However, both of these energy densities are considerably smaller than
that in thermal particles $\epsilon_{\rm B} (\sim \epsilon_{\rm rel}) \sim
1\times 10^{-3} (\dot{M}_{-5}/v_{\rm w,3})^{-1}$, unless the mass loss rate
of the progenitor is small.  The actual values of $U_{\rm B}$ and
$U_{\rm rel}$ deduced above are roughly the same as those derived by BKC
using radio data alone (in particular, the observed value of the
synchrotron self-absorption frequency) and {\it assuming}
equipartition. The fact that these two independent methods to determine
the energy densities in magnetic fields and relativistic electrons give
the same value lends support to the Compton cooling scenario.  This is
further strengthened by the determination of the shock velocity, for
which both methods give approximately the same value.

Internally produced synchrotron photons are unlikely to contribute to
the cooling as can be seen from the following argument. When cooling
is important and $p\approx 2.0$, the energy density of synchrotron
photons is $U_{\rm ph}\sim U_{\rm B}(v_{\rm sh}/c)(\epsilon_{\rm
rel}/\epsilon_{\rm B})^{1/2}$. The condition corresponding to equation
(\ref{eq:1.8}) then becomes $\epsilon_{\rm rel} (\dot{M}_{-5}/v_{\rm
w,3})>4.7 (v_{\rm sh}/c)^{-4/3}(\epsilon_{\rm rel}/\epsilon_{\rm
B})^{1/3}$.  Since $\epsilon_{\rm rel}/\epsilon_{\rm B} > 1$ is
necessary for Compton scattering to dominate synchrotron radiation,
this requires an even higher mass loss rate than the synchrotron
cooling scenario.

\section{A model fit to the radio and X-ray radiation of SN 2002ap}
\label{sec_3}
The estimates above show that Compton cooling can provide a natural
scenario for SN 2002ap. A more detailed model fit to the
observations is therefore warranted.  We have for this purpose used
the numerical model in \citet{FB98}, which solves the radiative
transfer equation for the synchrotron radiation, including
self-absorption, together with the kinetic equation for the electron
distribution, including synchrotron, Compton and Coulomb losses. The
latter is unimportant for SN 2002ap. As discussed above, we assume
that a constant fraction, $\epsilon_{\rm B}$, of the thermal energy
behind the shock goes into magnetic fields and a fraction,
$\epsilon_{\rm rel}$, into relativistic electrons. The other important
input parameters are $p$, $\Mdot/v_{\rm w}$, and $n$, the ejecta density
power law index (specifying the shock velocity, see above).  Finally,
the bolometric luminosity, $L_{\rm bol}$, determines the Compton
cooling. For $L_{\rm bol}$ we use the bolometric light curve
determined by \citet{Maz02}, \citet{YO03}, and \cite{P03}. Because free-free
absorption is unimportant in the cases we consider, and the
synchrotron self-absorption is determined by $B$, $n_{\rm rel}$ and $R$
only, $\Mdot/v_{\rm w}$ always enters in the combination
$\epsilon_{\rm B}\Mdot/v_{\rm w}$ and $\epsilon_{\rm rel} \Mdot/v_{\rm w}$,
reducing the number of free parameters by one, but also preventing us
from determining $\Mdot/v_{\rm w}$ separately.

\begin{figure}
\epsscale{0.8}
\plotone{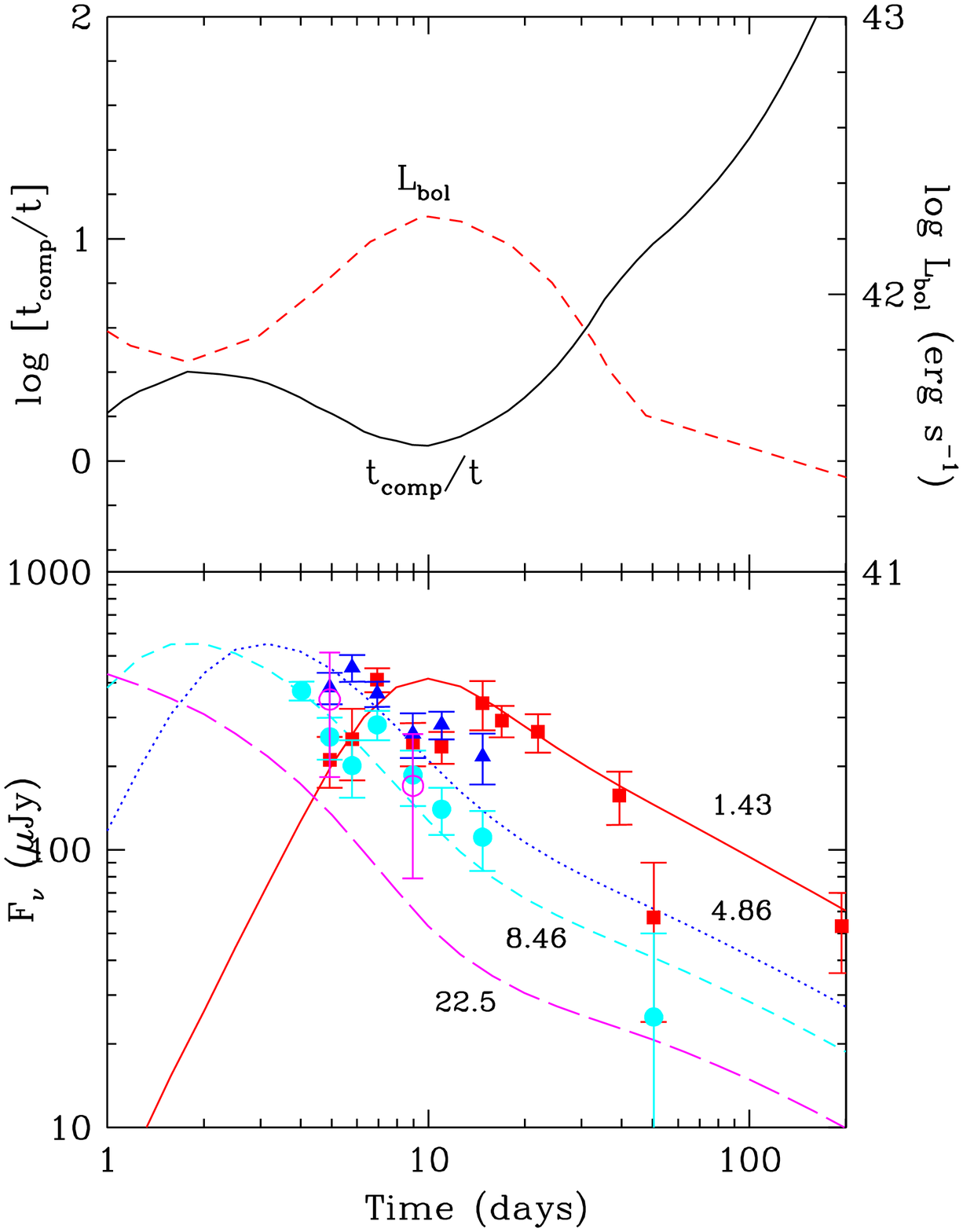}
\caption{Upper panel: The solid line shows the ratio of the Compton 
cooling time scale for electrons radiating at $22.5$ GHz to the
adiabatic time scale as function of time for the model in the lower
panel. The dashed line gives the bolometric luminosity taken from
\protect\citet{Maz02}, \protect\citet{YO03}, and \protect\citet{P03}. 
Lower panel: Radio light curves for SN 2002ap. Parameters are
given in the text. The observational VLA data are from BKC ($1.43$ 
GHz--{\it squares}; $4.86$ GHz--{\it triangles}; $8.46$ GHz--{\it 
filled circles}; $22.5$ GHz--{\it open circles}).}
\label{fig1}
\end{figure}

\begin{figure}
\rotatebox{-90}{
\epsscale{0.8}
\plotone{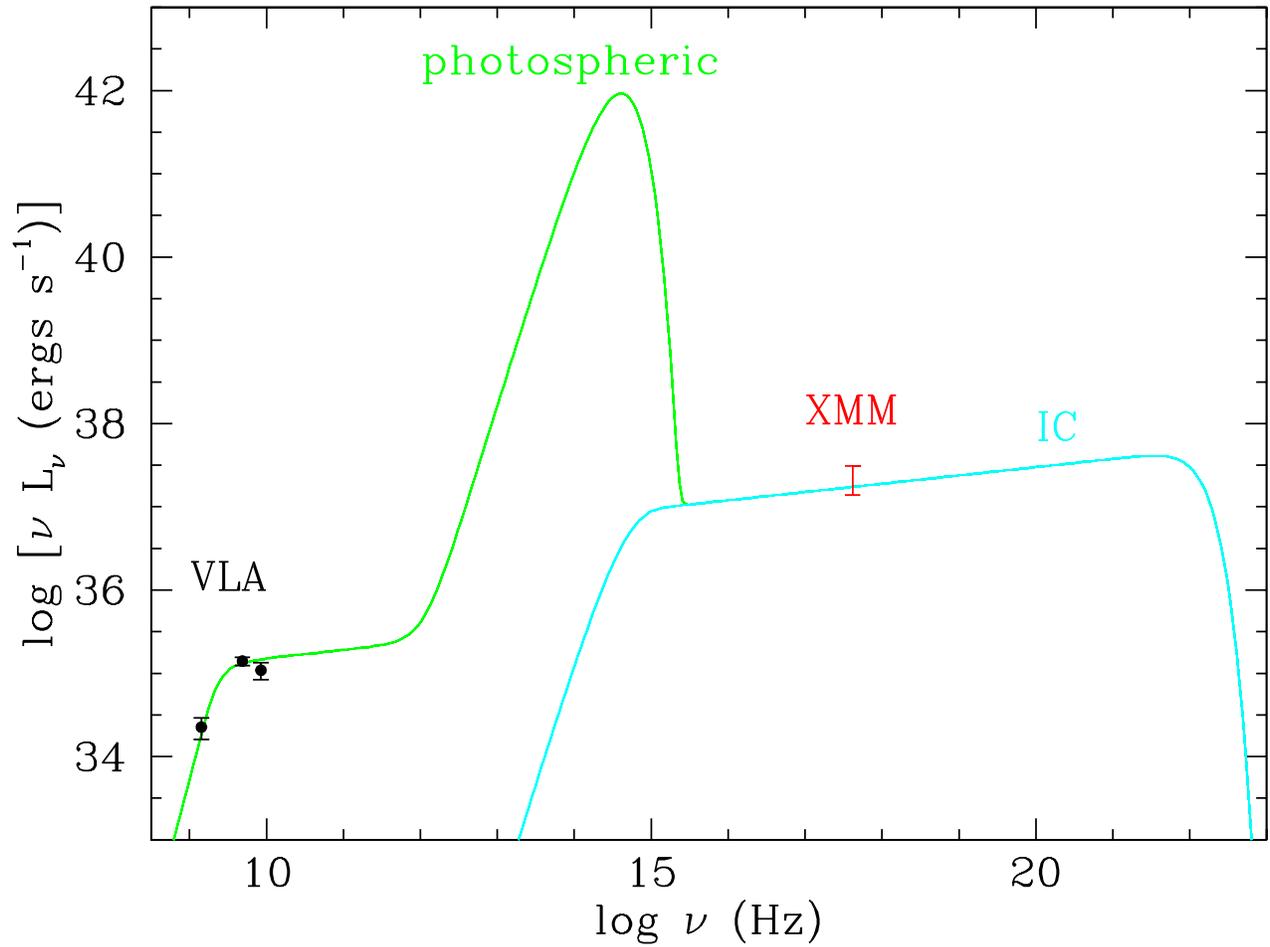}}
\caption{Spectrum ($\nu L_\nu$) of SN 2002ap 6 days after
  explosion. The radio fluxes are from BKC, while the X-ray flux is 
  derived from XMM-observations (see text for a discussion of the 
  uncertainties associated with this value).}
\label{fig2}
\end{figure}

Using this model, we have varied the parameters to give a best fit of
the radio light curves together with the XMM flux at 6 days, taken
from \citet{SCB03}. In Figure \ref{fig1} we show the resulting light
curves and in Figure \ref{fig2} the full spectrum at 6 days, together
with the VLA and XMM observations. In this figure we have also added
the luminosity from the supernova photosphere. The latter can on day 6
be well approximated as a black-body with temperature of $\sim 5000$ K
and a total luminosity of $1.6\EE{42}\ergs$ \citep{P03}.  As input
electron spectrum we use $p=2.1$, and $\gamma_{\rm min}=1$ and
$\gamma_{\rm max}=5\EE3$. The latter is only important for the upper
energy limit of the gamma-ray flux from inverse Compton.

In order to reproduce the decrease in the optically thin parts of the light
curves, the best fit model has $n=10$, in agreement with the expected
density structure for the progenitor of an Ic SN \citep{MMK99}. 
This is also consistent with hydrodynamical models of Type Ic
supernovae, where the power law index varies between $n=8$ and $n=13$ 
for ejecta velocities above $\sim 25,000 \kms$ \citep{Iwa00}.The
expansion velocity we find is $70,000\kms$ at 10 days, somewhat lower
than deduced by BKC. As discussed in \S \ref{sec_2.2}, the requirement
that inverse Compton scattering should reproduce both the X-ray flux
and the synchrotron radio spectrum, sets the value of $\epsilon_{\rm
B}/\epsilon_{\rm rel}$ as well as $\epsilon_{\rm B} \Mdot/v_{\rm w}$,
and we find that $\epsilon_{\rm B}/\epsilon_{\rm rel} \approx 0.2$ and
$\epsilon_{\rm B} \Mdot/v_{\rm w}\approx 1\EE{-11} \ml / \kms$. Note 
that $\epsilon_{\rm rel}$ is roughly proportional to 
$\ln(\gamma_{\rm max}/\gamma_{\rm min})$. Since the values of both 
$\gamma_{\rm min}$ and $\gamma_{\rm max}$ are unknown, this introduces an 
uncertainty in the value of $\epsilon_{\rm rel}$; for example, using 
$\gamma_{\rm min}=\gamma_{\rm abs}$, where $\gamma_{\rm abs}$ is the 
Lorentz factor of the electrons radiating at the synchrotron 
self-absorption frequency, yields $\epsilon_{\rm B}/\epsilon_{\rm 
rel}\approx 1$. As
anticipated in \S \ref{sec_2.2}, inverse Compton cooling of the
electrons is important, leading to a steepening of the electron
spectrum and a spectral flux $F_\nu\propto \nu^{-0.9}$, in
agreement with observations (BKC). The importance of
Compton cooling for the calculated light curves can be seen especially 
at high
frequencies, where the cooling time is comparable to the dynamical 
time (cf. Fig. \ref{fig1}). For these a depression is apparent at the
time of maximum Compton cooling at $\approx 10$ days.

\section{Discussion}
\label{sec_disc}  
In order to derive the main parameters characterizing an observed
synchrotron source of unknown size, one often has to invoke the assumption of
equipartition between energies in relativistic electrons and magnetic
field (i.e., $\epsilon_{\rm B} \approx \epsilon_{\rm rel}$). The
independent determination of these quantities requires either that
radiative cooling is important within the observed frequency range, or
that the Compton scattered radiation is observed. In the latter case,
the scattered radiation can be the synchrotron photons themself or
come from a known external source of photons.

We have shown in this paper that the radio, optical and X-ray
observations of SN 2002ap can be understood in a scenario, wherein the
relativistic electrons cool by inverse Compton scattering on the
photospheric supernova photons and thereby giving rise to the X-ray
emission. Since both radiative cooling and the Compton scattered
emission are observed, not only can a value for $\epsilon_{\rm B} /
\epsilon_{\rm rel}$ be derived but, in addition, an internal
consistency check of the model can be made; for example, the value of
the synchrotron self-absorption frequency can be predicted. The
Compton cooling scenario {\it implies} rough equipartition conditions in SN
2002ap. Furthermore, the actual value deduced for $\epsilon_{\rm B}$
($\sim \epsilon_{\rm rel}$) is close to the one derived by BKC from
the observed values of the synchrotron self-absorption frequency and
flux {\it assuming} equipartition. This shows that the model
self-consistently predicts the frequency of the synchrotron
self-absorption, as was illustrated in \S \ref{sec_3}.

There is, however, one underlying assumption; namely, a spherically
symmetric source. It is seen in \S \ref{sec_2.2} that the only place
where the assumption of a spherically symmetric source geometry enters
is in the derivation of the energy density of relativistic electrons
from the observed X-ray luminosity (cf. eq. [\ref{eq:1.18}]).
Deviations from spherical symmetry, for example a jet structure, would
then have to be compensated for by an increased energy density of
relativistic electrons. Since the deduced values of $B$ and $v_{\rm
sh}$ are not affected (cf. eqs. [\ref{eq:1.16}] and[\ref{eq:1.17}]), this
results in an increased synchrotron self-absorption frequency.  Due to
the short term flux variations, which are likely caused by
interstellar scattering and scintillation, the limits of the allowed
variations of the predicted self-absorption frequency are hard to evaluate
precisely. From the early observations at the lowest frequency (1.43
GHz), when the radiation at this frequency was optically thick,
it seems unlikely that the observed self-absorption frequency
has been underestimated by more than a factor two. Since synchrotron
self-absorption frequency scales with energy density of relativistic
electrons as $\epsilon_{\rm rel}^{2/(p+4)}$, the value of
$\epsilon_{\rm rel}$ can be increased at most by a factor ten. Hence,
the solid angle of a jet has to cover at least 10~\% of the sky. This
conclusion is similar to the one reached by \citet{Tot03}, using a
different line of reasoning.

BKC argue that the observed XMM flux can be described as an
extrapolation of the synchrotron flux. This is apparently based on the
assumption that the radio spectral index is close to $\beta\sim 0.5$,
which is needed to explain the light curves in the absence of
cooling. The resulting spectral break in the optical frequency range 
due to synchrotron cooling in an
equipartition magnetic field would give rise to an X-ray flux and
spectral index in approximate agreement with those observed. However,
as discussed in \S \ref{sec_2.2}, the observed spectral index in the
radio is $\beta\sim 0.9$, which is similar to that in the X-ray range,
making this scenario untenable.

\citet{SCB03} explain the X-ray flux as a result of {\it thermal}
inverse Compton scattering by the thermal electrons behind the shock
\citep[see][]{F82}. In order to have sufficient electron optical depth,
they need to have the X-ray emitting region moving with a velocity 
$\sim 16,000\kms$. A
similar low velocity of the forward shock is needed in the model by
\citet{SPM03}. They argue that the X-ray emission is free-free
emission from the reverse shock. There are several features of these
models which make them less attractive. In the thermal
inverse Compton scattering model, the spectral index is very sensitive
both to shock temperature and optical depth. The low shock 
temperature in the free-free
model requires a density structure of the ejecta corresponding to
$n\sim 40$ at velocities $\sim 10,000-20,000\kms$, which is quite 
different from that thought appropriate fora Ic SN, $n\la 10$ 
\citep{Iwa00,MMK99}, in this velocity range. 

In addition to these model specific problems, there are two rather
model independent arguments against any model invoking such low shock
velocities. First, the line emission observed in SN 2002ap should set
a lower limit to the velocity of the forward shock. Due to line
blending, accurate velocities were hard to measure in SN 2002ap. However, it seems
clear that velocities of at least $0.1~\rm c$ are indicated. \cite{Maz02} 
find in their first spectrum at 2 days a photospheric
velocity of 30,000 km/s, and at 3.5 days 20,500 km/s.  It is important
to note that this is the velocity where the {\it continuum} is becoming
optically thick and, therfore, provides a lower limit to the velocity 
of the {\it line} emitting regions. This is confirmed by the
observations of \citet{Fol03}, which show that the P-Cygni profile of
the O I $\wl 7774$ line on day 15 extends at least to $\sim
24,000\kms$, where it becomes blended with other lines. This, in turn, should be
a conservative lower limit to the {\it shock} velocity and, hence, calls in
question the consistency of the proposed models.

Another issue for low velocity models is the incorporation of
the observed radio emission.  Assuming the radio emission to come from
behind the forward and/or the reverse shock, implies a brightness
temperature which scales, roughly, as $v_{\rm sh}^{-2}$. The low shock
velocities argued for in the above two models result in brightness
temperatures about a factor 10 larger than the Compton
limit. This, in turn, would produce an X-ray flux much in excess to
that observed.

The rough equipartion between the energies in relativistic electrons 
and magnetic fields found for SN 2002ap is in sharp contrast to the 
conditions in SN 1993J for which $\epsilon_{\rm B}/\epsilon_{\rm rel}\gg 
1$ was deduced by \citet{FB98}. Unfortunately, these are the only two 
SNe for which an independent estimate of $\epsilon_{\rm B}/\epsilon_{\rm rel}$ 
has been possible to make. In this context it 
is interesting to note that for the afterglows of GRBs the deduced values of 
$\epsilon_{\rm B}/\epsilon_{\rm rel}$ exhibit a large range 
\citep[e.g.,][]{PK02}. However, there is a clear tendency for this ratio to be 
smaller, or for some afterglows much smaller, than unity. Although 
the conditions for both magnetic field generation and particle 
acceleration may differ considerably between relativistic and 
non-relativistic shocks, a plausible  cause for the substantial difference in 
the value of $\epsilon_{\rm B}/\epsilon_{\rm rel}$ behind the non-relativistic 
shocks in SN 1993J and SN 2002ap is harder to find. 

\section{Conclusions}
\label{sec_concl}  
We have shown that in SN 2002ap inverse Compton scattering of the
photospheric photons is important for those relativistic electrons
producing the radio emission. The scattering cools the relativistic
electrons, and gives rise to an X-ray flux which matches the observed
flux level as well as the spectral index. The value for the shock velocity, 
$\sim 70,000 \kms$,
is in rough agreement with BKC. The deduced energy densities
in magnetic fields and relativistic electrons are close to
equipartition. However, without {\it a priori}
knowledge of what fraction of the total injected energy density these
quantities correspond to, no estimate can be made of the mass
loss rate of the progenitor star. From the radio and X-ray
observations the mass loss rate can be constrained to
$\Mdot \approx 1\EE{-8} (v_{\rm w}/1000 \kms)\epsilon_{\rm B}^{-1}
\ml$.  BKC assumed $\epsilon_{\rm
B}=\epsilon_{\rm rel} = 0.1$, which would imply $\Mdot \approx
1\EE{-7} (v_{\rm w}/1000 \kms) \ml$; this is very low for a
Wolf-Rayet star. Turning this argument around, a typical Wolf-Rayet
mass loss rate of $\Mdot \sim 10^{-5}\ml$ and wind velocity of $\sim
2000 \kms$, would imply $\epsilon_{\rm B}\sim 2\EE{-3}$.

An important conclusion from this paper is that any self-consistent
model of the radio and X-ray flux observed from SN 2002ap needs to
include effects due to the photospheric emission. It is likely that in
any SNe, for which the synchrotron emission from behind the shock
becomes transparent not too long after the photospheric emission has
reached its maximum, inverse Compton scattering can be an important
factor shaping the radio and/or X-ray spectrum. Another outcome of our
analysis is that the time variation of the radio emission allows a
determination of the density structure 
of the ejecta. It is reassuring that the result is close to what is 
expected theoretically.

\acknowledgements

This research was supported by grants from the Swedish 
Science Research Council. We are grateful for the comments and 
suggestions provided by Roger Chevalier.

\clearpage

\begin{thebibliography}{}

\bibitem[Berger, Kulkarni, \& Chevalier(2002)]{BKC02} Berger, 
E., Kulkarni, S.~R., \& Chevalier, R.~A.\ 2002, \apjl, 577, L5 

\bibitem[Chevalier(1982)]{RAC82} Chevalier, R.~A.\ 1982, 
\apj, 258, 790 

\bibitem[Chevalier \& Fransson(2003)]{CF03} Chevalier, R.~A.~\&
Fransson, C.\ 2003, in "Supernovae and Gamma-Ray Bursts," edited by
K. W. Weiler (Springer-Verlag), 171

\bibitem[Chevalier \& Li(2000)]{CL00} Chevalier, R.~A.~\& 
Li, Z.\ 2000, \apj, 536, 195

\bibitem[Foley et al.(2003)]{Fol03} Foley, R.~J.~et 
al.\ 2003, \pasp, 115, 1220 

\bibitem[Fransson(1982)]{F82} Fransson, C.\ 1982, \aap, 
111, 140 

\bibitem[Fransson \& Bj{\" o}rnsson(1998)]{FB98} Fransson, 
C.~\& Bj{\" o}rnsson, C.-I.\ 1998, \apj, 509, 861 

\bibitem[Hjorth et al.(2003)]{Hjo03} Hjorth, J., ~et al.\ 
2003, \nat, 423, 847

\bibitem[Iwamoto et al.(2000)]{Iwa00} Iwamoto, K.~et al.\ 
2000, \apj, 534, 660

\bibitem[Iwamoto et al.(2003)]{Iwa03} Iwamoto, K., Nomoto, K.,
Mazzali, P.~A., Nakamura, T., \& Maeda, K.\ 2003, in "Supernovae and
Gamma-Ray Bursts," edited by K. W. Weiler (Springer-Verlag), 243
 
\bibitem[Kawabata et al.(2003)]{Kaw03} Kawabata, K.~S.~et 
al.\ 2003, \apjl, 593, L19 

\bibitem[Li \& Chevalier(2003)]{LC03} Li, Z.~\& Chevalier, 
R.~A.\ 2003,  in "Supernovae and Gamma-Ray Bursts," edited by
K. W. Weiler (Springer-Verlag), 419

\bibitem[Lundqvist \& Fransson(1988)]{LF88} Lundqvist, P.~\& 
Fransson, C.\ 1988, \aap, 192, 221 

\bibitem[Matzner \& McKee(1999)]{MMK99} Matzner, C.~D.~\& 
McKee, C.~F.\ 1999, \apj, 510, 379 

\bibitem[Mazzali et al.(2002)]{Maz02} Mazzali, P.~A.~et al.\ 
2002, \apjl, 572, L61 

\bibitem[Nomoto et al.(1994)]{N94} Nomoto, K., Yamaoka, H., 
Pols, O.~R., van den Heuvel, E.~P.~J., Iwamoto, K., Kumagai, S., \& 
Shigeyama, T.\ 1994, \nat, 371, 227 

\bibitem[Panaitescu \& Kumar(2002)]{PK02} Panaitescu, A.~\& 
Kumar, P.\ 2002, \apjl, 571, 779 

\bibitem[Pandey et al.(2003)]{P03} Pandey, S.~B., Anupama, 
G.~C., Sagar, R., Bhattacharya, D., Sahu, D.~K., \& Pandey, J.~C.\ 2003, 
\mnras, 340, 375 

\bibitem[Sharina, Karachentsev, \& Tikhonov(1996)]{SKT96} 
Sharina, M.~E., Karachentsev, I.~D., \& Tikhonov, N.~A.\ 1996, \aaps, 119, 
499 

\bibitem[Soria \& Kong(2002)]{SK02} Soria, R.~\& Kong, 
A.~K.~H.\ 2002, \apjl, 572, L33 
 
\bibitem[Soria, Pian, \& Mazzali(2004)]{SPM03} Soria, R., 
Pian, E., \& Mazzali, P.~A.\ 2004, \aap, 413, 107 

\bibitem[Stanek et al.(2003)]{Stan03} Stanek, K.~Z.~et al.\ 
2003, \apjl, 591, L 17

\bibitem[Sutaria, Chandra, Bhatnagar, \& Ray(2003)]{SCB03} 
Sutaria, F.~K., Chandra, P., Bhatnagar, S., \& Ray, A.\ 2003, \aap, 397, 
1011 

\bibitem[Totani(2003)]{Tot03} Totani, T.\ 2003, \apj, 598, 
1151 

\bibitem[Woosley, Langer, \& Weaver(1995)]{WLW95} Woosley, 
S.~E., Langer, N., \& Weaver, T.~A.\ 1995, \apj, 448, 315 

\bibitem[Yoshii et al.(2003)]{YO03} Yoshii, Y.~et al.\ 2003, 
\apj, 592, 467 



\end{thebibliography}
\end{document}